# Athermal jamming of soft frictionless Platonic solids

Kyle C. Smith,[1] Meheboob Alam,[2] and Timothy S. Fisher[1]*

[1] *School of Mechanical Engineering, Birck Nanotechnology Center, and Maurice Zucrow Laboratories, Purdue University, West Lafayette, IN 47907, USA*

[2] *Engineering Mechanics Unit, Jawaharlal Nehru Centre for Advanced Scientific Research, Jakkur Campus, Bangalore 560064, India*

*corresponding author's email: *tsfisher@purdue.edu*

**Abstract**

A mechanically-based structural optimization method is utilized to explore the phenomena of jamming for assemblies of frictionless Platonic solids. Systems of these regular convex polyhedra exhibit mechanically stable phases with density substantially less than optimal for a given shape, revealing that thermal motion is necessary to access high density phases. We confirm that the large system jamming threshold of 0.623 ± 0.003 for tetrahedra is consistent with experiments on tetrahedral dice. Also, the extremely short-ranged translational correlations of packed tetrahedra observed in experiments are confirmed here, in contrast with those of thermally simulated glasses. Though highly ordered phases are observed to form for small numbers of cubes and dodecahedra, the short correlation length scale suppresses ordering in large systems, resulting in packings that are mechanically consistent with 'orientationally disordered' contacts (point-face and edge-edge contacts). Mild nematic ordering is observed for large systems of cubes, whereas angular correlations for the remaining shapes are ultra short-ranged. In particular the angular correlation function of tetrahedra agrees with that recently observed experimentally for tetrahedral dice. Power-law scaling exponents for energy with respect to distance from the jamming threshold exhibit a clear dependence on the 'highest order' percolating contact topology. These nominal exponents are 6, 4, and 2 for configurations having percolating point-face (or edge-edge), edge-face, and face-face contacts, respectively. Jamming contact number is approximated for small systems of tetrahedra, icosahedra, dodecahedra, and octahedra with order and packing representative of larger systems. These Platonic solids exhibit hypostatic behavior, with average jamming contact number between the isostatic value for spheres and that of asymmetric particles. These shapes violate the isostatic conjecture, displaying contact number that decreases monotonically with sphericity. The common symmetry of dual polyhedra results in local translational structural similarity. Systems of highly spherical particles possessing icosahedral symmetry, such as icosahedra or dodecahedra, exhibit structural behavior similar to spheres, including jamming contact number and radial distribution function. These results suggest that though continuous rotational symmetry is broken by icosahedra and dodecahedra, the structural features of disordered packings of these particles are well replicated by spheres. Octahedra and cubes, which possess octahedral symmetry, exhibit similar local translational ordering, despite exhibiting strong differences in nematic ordering. In general, the structural features of systems with tetrahedra, octahedra, and cubes differ significantly from those of sphere packings.

**I.     Introduction**

Granular materials are observed in nature and widely utilized in modern industrial processes. Aside from materials possessing solid granular particles, other discrete network systems such as foams, emulsions, and glasses display similar mechanical and structural behavior [1]. The mechanics of granular materials can exhibit gas, liquid, and solid-like behavior [2]. Density, defined as the fraction of volume occupied by grains, and state of stress are critical quantities of interest for such systems of particles. For periodic systems, density $\phi$ is defined as $NV_p/V_{cell}$, where $N$, $V_p$, and $V_{cell}$ are the number of particles in the primary cell, volume of each particle, and volume of the primary cell, respectively. In particular, for systems of particles with purely repulsive interactions the jamming threshold density $\phi_J$ is of interest. The jamming threshold marks the thermodynamic transition between flowing and static states of granular materials [3]. In practice the onset of non-zero potential energy has been used to identify the jamming threshold density (*e.g.*, in [3]).

In addition to the jamming threshold, which could potentially depend on consolidation path, the optimal density for a given particle shape is also an important quantitative characteristic of granular assemblies. Optimal density differs from jamming threshold in that optimal density is the maximum possible density for an assembly of particles irrespective of its accessibility via mechanical or thermal means. A proof of optimal density for spheres was not accomplished until recently [4]. Recent non-mechanically-based approaches have been employed to determine the optimal density of Platonic and Archimedean solids [5-6], tetrahedra [5-13], ellipsoids [14], and superballs [15]. Both geometric [7, 9-13] and thermodynamic Monte Carlo [5, 6, 8] methods have been utilized to probe high density phases of rigid particles, but incremental random motions that particles undergo in such simulations are not induced by mechanical interaction. As a result, such methods enable systems to bypass low density jammed configurations. Accordingly, these methods are capable of achieving densities and phases not obtainable through purely athermal mechanical interactions. In particular, several groups have focused intensely on demonstrating increases in the predicted maximum density of tetrahedra packings, employing Monte Carlo and geometric methods. A Monte Carlo method was employed to find a densest packing of $\phi$ = 0.823 [5], exceeding previous record densities obtained through strictly geometric approaches (see [10-12]). Haji-Akbari *et al.* [8] reported the spontaneous formation of a quasi-crystal for a fluid of tetrahedra at $\phi$ = 0.8230, and by compressing an approximant of that structure they obtained a structure with $\phi$ = 0.8503. Kallus *et al.* [13] utilized a 'divide and conquer' search of non-overlapping positions and orientations of four tetrahedra, and thereby discovered a one-parameter family of dense double-dimer packings with $\phi$ = 0.8547. Chen *et al.* [9] extended the double-dimer system to produce the highest density structure of tetrahedra with $\phi$ = 0.856347, which was subsequently shown to be the optimal packing for a six-parameter family of double-dimers [7]. These densities are much higher than the experimentally reported value of $\phi$ = 0.76 ± 0.02 for tetrahedral dice [16]. Further, Jiao and Torquato [5, 6] showed that optimal packings of icosahedra, dodecahedra, and octahedra are crystalline using Monte Carlo optimization, and they can form structures with $\phi$ > 0.8.

Though Monte Carlo and geometric approaches are useful for finding optimal packings of particles, many granular systems cannot access the highest density phases. For example, although spheres possess an optimal closest packed density of $\phi$ = 0.74, the maximally random jammed state of spheres is known to produce a density of $\phi$ = 0.64 [17]. Thus, to obtain granular configurations that are commonly encountered in nature and industry the methods utilized to consolidate granular materials should be replicated. Athermal, mechanically-based studies of spheres [3] and ellipses [18] have been performed recently through energy based approaches and have yielded sphere packing densities consistent with the precisely-defined maximally random jammed density (see [17]). Though these methods do not incorporate friction, they do replicate the quasi-static nature of motion during consolidation.

Properties other than jamming threshold are of interest at the jamming point, as well. For instance, the average number of contacts per particle at the jamming point exhibits values that depend on particle shape. The isostatic conjecture attempts to describe structure at the jamming point through kinematic constraint counting. The conjecture suggests that the average number of contacts between particles is equal to twice the average number of particle degrees of freedom. Ellipses [18, 19], ellipsoids [19], and tetrahedra [16] for instance display hypostatic behavior, *i.e.* having average jamming contact numbers less than their respective isostatic values. The departure of ellipsoids from isostatic behavior has been attributed to the presence of floppy modes, which provide vanishing restoring force [18], whereas for tetrahedra it has been attributed to the varying degrees of rotational constraint by discrete contact topologies (*e.g.*, vertex-face, edge-edge, edge-face, face-face contacts) [16].

Mechanical properties of granular systems as they are perturbed to densities above the jamming threshold have exhibited power law scaling [3]. Expressed mathematically, properties (*e.g.*, system energy, pressure, contact number) denoted as $X$ vary with density $\phi$ relative to the jammed density $\phi_J$ as:

$$X \sim \left(\phi - \phi_J\right)^{\beta} \tag{1}$$

Scaling exponents for mechanical properties $\beta$ have not been shown to be universal. For instance, bulk and shear modulus scaling exponents depend on the contact force model employed in simulations. O'Hern *et al.* [3] utilized various potential interaction models between bi-disperse disks in which potential energy scaled with inter-particle separation via power-law exponents of 5/2 (Hertzian) and 2 (harmonic), respectively. For the different potential models utilized by O'Hern *et al.*, bulk modulus displayed scaling exponents of 1/2 and zero, while shear modulus displayed scaling of 1 and 1/2, respectively. In contrast to scaling of mechanical properties the excess contact number, $Z$-$Z_J$, exhibited 1/2 power-law scaling for a variety of contact interaction models [3]. Ellenbroek *et al.* [20] explained this dependence by relating the probability of developing a new contact with the scaling of non-affine displacements away from the jamming point. Universality of this scaling across particle morphologies is not guaranteed though.

In this work, we explore the jamming of assemblies of soft Platonic solids (Fig. 1) through mechanically-based structural optimization. Particular emphasis in Section II is given to the methods of structural optimization and the calculation of forces on particles. In Section III results are presented for the distribution of jamming threshold for small configurations of particles, scaling of mechanical properties upon perturbation from the jamming point, system ordering, and isostaticity.

**II.   Methodology**

At the onset of jamming during consolidation, total energy is infinitesimal and any motion of a collection of particles necessarily increases total energy. Approaching this point is quite difficult because the finite range of potentials employed in soft particle simulations requires finite energy to impart forces between particles to induce motion. Consequently small perturbations from the jamming point, having finite, non-zero total energy are required for numerical simulation. To obtain such configurations, a sequence of compressive and expansive volumetric strain steps is applied to an initially dilute ($\phi$ = 0.05) three-dimensional configuration of $N$ particles in the primary cubic-shaped cell. Periodic boundary conditions are often employed to neglect wall effects in jamming simulations, as reflected in the recent review on jamming of hard particles by Stillinger and Torquato [21]. Accordingly, we employ periodic boundary conditions for all configurations studied in this work. Following each strain step, conjugate gradient energy minimization is utilized to relax the system of particles toward static equilibrium. The conjugate gradient method admits simultaneous motion of particles, resulting in equilibrium configurations that are collectively jammed upon convergence, *i.e.* energy of these configurations cannot be reduced by individual or concerted particle motions [23].

Here the initial dilute configuration is chosen with pseudo-random particle positions and orientations. In contrast to the work of O'Hern *et al.* [3] to find inherent structures of particle systems, our approach attempts to replicate the stepwise consolidation of random particle assemblies to dense, jammed states. Configurations are compressed past the jamming point near a target energy $E_t$ and subsequently expanded toward the jamming point. The initial configuration is first compressed in an affine manner by an isotropic volumetric strain increment $\varepsilon_V$. Conjugate gradient minimization then proceeds until (1) average energy $E$ falls below a threshold value of $0.25E_t$ or (2) equilibrium is achieved. Equilibrium convergence criterion is described in sub-section B. When criterion (1) is satisfied ($E < 0.25E_t$) the configuration is further compressed by $\varepsilon_V$ to approach a mechanically stable state near the consolidation target energy. Above the jamming point, criterion (2) will be satisfied, and if energy is less than the target energy ($E < E_t$), the jammed configuration possesses acceptable energy. In contrast, if the equilibrium energy is greater than the target energy ($E > E_t$) the configuration is expanded by $-0.75\varepsilon_V$. Following expansion, $\varepsilon_V$ is recursively assigned a value of $0.25\varepsilon_V$ in order to asymptotically approach the target energy with further strain-relaxation sequences. This procedure is similar to that of Mailman *et al.* [18] and Gao *et al.* [22] in which volumetric strain is applied and subsequently relaxed via conjugate gradient iterations.

Following consolidation, the assembly is sequentially expanded to approach the jamming point ($E \rightarrow 0^+$). To approach the jamming point asymptotically, the strain increment is chosen to depend on the current configuration's energy $E$ and the contact model exponent $m$:

$$\varepsilon_V = \gamma E^{1/(3m)} \qquad (2)$$

Here, $\gamma$ is a constant chosen for a particular system of particles to achieve sufficient system energy variation during the expansion process. As will be shown later, this strain is generally insufficient to reduce energy directly to zero for disordered packings, and will result in an asymptotic approach toward the jamming point.

## A. Conjugate Gradient Method

The conjugate gradient method is employed to search for a local minimizer of the positions and orientations of particles in the configuration. The conjugate gradient method is implemented by considering the energy gradient $\mathbf{g}_i$ for each particle $i$:

$$\mathbf{g}_i = -\mathbf{S}_i^{-1/2} \begin{bmatrix} \mathbf{F}_i^T & \mathbf{M}_i^T \end{bmatrix}^T \qquad (3)$$

in terms of the net force $\mathbf{F}_i$ and moment $\mathbf{M}_i$ about its center of mass. $\mathbf{S}_i$ is a block diagonal matrix corresponding to particle $i$ in the assembly:

$$\mathbf{S}_i = \begin{bmatrix} \mathbf{I}_3 & 0 \\ 0 & D_i^2 \mathbf{I}_3 \end{bmatrix} \qquad (4)$$

Here $\mathbf{I}_3$ is the 3x3 identity matrix, and $D_i$ is the diameter of the bounding sphere coincident with the particle's centroid. The gradient vector for the system $\mathbf{g}$ is a row-wise concatenation of those for each particle $\mathbf{g}_i$. The conjugate gradient minimization procedure starts with a trajectory in the steepest descent direction [27]:

$$\mathbf{r}^{(1)} = -\mathbf{g}^{(1)} \tag{5}$$

The elements of $\mathbf{r}^{(k)}$ are actually displacement coordinates in real-space and must be converted to angular-space to perform rotation operations on particles. The properly scaled search trajectory $\mathbf{s}^{(k)}$ is given by:

$$\mathbf{s}^{(k)} = \mathbf{S}^{-1/2} \mathbf{r}^{(k)} \tag{6}$$

where $\mathbf{S}$ is a block-diagonal matrix incorporating $\mathbf{S}_i$ sub-matrices intended to scale rotational coordinates relative to translation coordinates:

$$\mathbf{S} = \begin{bmatrix} \mathbf{S}_1 & 0 & 0 \\ 0 & \cdot & 0 \\ 0 & 0 & \mathbf{S}_N \end{bmatrix} \tag{7}$$

Translation and rotation components for each particle are decomposed from $\mathbf{s}^{(k)}$ and are then applied to particles in order to compute energy at the corresponding position. Details of the quaternion-based rotation operation utilized are described in Appendix A.

  A one-dimensional line search for the minimum along this direction is performed. Details of this line search procedure employed are described in the Appendix B. The trajectory for the next conjugate gradient search is determined with the following update formula

$$\mathbf{r}^{(k+1)} = -\mathbf{g}^{(k+1)} + \beta^{(k)} \mathbf{r}^{(k)} \tag{8}$$

where $\beta^{(k)}$ is determined by the formula of Polak and Ribiere [28]:

$$\beta^{(k)} = \left(\mathbf{g}^{(k+1)} - \mathbf{g}^{(k)}\right)^T \mathbf{g}^{(k+1)} \Big/ \left(\left(\mathbf{g}^{(k)}\right)^T \mathbf{g}^{(k)}\right) \qquad (9)$$

The formula of Fletcher and Reeves [29] was also tested and found to substantially underperform. Though both formulas produce quadratic rates of convergence when applied to a quadratic objective functional, the Fletcher-Reeves formula can slow to linear convergence for non-quadratic objective functionals if not sequentially restarted (see [30]). The Polak-Ribiere update formula tends to restart automatically toward the steepest descent direction ($\mathbf{r}^{(k+1)} \to -\mathbf{g}^{(k+1)}$ when $\mathbf{g}^{(k+1)} \to \mathbf{g}^{(k)}$), and as a result it is superior for the minimization of energy functionals employed in this work.

Convergence of conjugate gradient iterations is assessed by considering the relative change in energy $D_E^{(k)}$ at iteration $k$:

$$D_E^{(k)} = \frac{E^{(k)} - E^{(k-1)}}{E^{(k-1)}} \qquad (10)$$

This convergence criterion has been utilized by Mailman *et al.* [18] and Gao *et al.* [22] to obtain mechanically stable packings of ellipsoids and disks, respectively, and we confirm that when set to an adequately small value, residual forces also become small relative to contact forces, satisfying static equilibrium.

### B. Contact energy model

Determining an expression for energy and force between contacting polyhedral particles in terms of inter-particle separation, as is done for spheres (*cf.* [3]) and ellipsoids (*cf.* [18]), is difficult due to the non-smooth surfaces of faceted particles. We utilize a simplified model proposed by Feng and Owen [31] whose energy-based contact model assumes that the elastic energy resulting from contact between two particles depends only on the volume of intersection

*V*. Variations of this model have been presented in two dimensions for discrete contact interaction cases by Pöschel and Schwager [32] and Feng and Owen [33]. Similar models have been applied in the literature wherein contact forces and moments are formulated heuristically in terms of contact geometry [34], but they lack energy conservation [32]. In contrast, the derivation of forces and moments from an energy basis results in energy conservation, though particle interpenetration may lead to unphysical behavior [32]. For the athermal, quasi-static simulations employed here, interpenetrating contacts are not expected because our implementation explicitly enforces that energy descent occurs during conjugate gradient iterations, preventing configurations from settling into local maxima during interpenetration.

The conjugate gradient method requires evaluation of contact forces and moments to determine search directions. Expressions for the forces and moments between a pair of particles, A and B, are hereafter presented. The force of particle A on particle B $\mathbf{F}_{AB}$ is the gradient of energy $E_{AB}$ with respect to the translational displacement of particle A $\mathbf{x}_A$.

$$\mathbf{F}_{AB} = \frac{dE_{AB}}{dV} \nabla_{\mathbf{x}_A} V \tag{11}$$

where *V* is the intersection volume between particles A and B. As indicated above in Eq. 11, the dependence of energy on volume (*i.e.*, $E_{AB}(V)$) influences the magnitude of force, but not its direction. The force's direction is determined entirely by the gradient of intersection volume. Intersection volume gradient is the sum of the product of area $A_i$ and inward pointing unit normal vector $\hat{n}_i$ for each face belonging to particle A on the intersection volume [31]:

$$\nabla_{\bar{x}_A} V = \sum_i A_i \hat{n}_i \tag{12}$$

Similarly, the moment about the centroid of particle B can be expressed as the sum from each face belonging to particle A on the intersection volume [31]:

$$\mathbf{M}_{AB} = \frac{dE_{AB}}{dV} \sum_i \mathbf{r}_i \times A_i \hat{n}_i \qquad (13)$$

where $\mathbf{r}_i$ is the distance from the centroid of particle A to the centroid of face $i$.

The form of the energy functional is important in representing the true dynamical character of the contacting particles, as well as in achieving stable numerical solutions. We assume that the energy functional takes the general form of a power law with exponent $m$, elastic modulus $Y$, and volumes of the particles in contact, $V_A$ and $V_B$:

$$E_{AB} = YV^m \left(V_A + V_B\right)^{1-m} / m \qquad (14)$$

The value of $m = 2$ utilized here ensures consistency with Hooke's law for mechanical interaction exhibited in uniaxial compression of aligned bars. Numerical details of the computation of intersection volume between contacting polyhedra are described in the Appendix C.

### III. Results and discussion

A sample curve displaying the variation of energy with density during the process described above is shown in Fig. 2. During the consolidation phase, system energy is maintained just below the target energy by performing conjugate gradient relaxation. At the final consolidation point, the system becomes mechanically stable near the target energy and is subsequently expanded through stable states toward the jamming point. We extrapolate the average contact depth with respect to density in order to approximate jamming threshold, as described in Appendix D. For each conjugate gradient iteration $D_E^{(k)} = 10^{-12}$ is used to assess convergence of 25 particle systems.

Choice of the target energy can influence the jamming threshold for a given initial configuration. Consolidated configurations of each shape were generated with $E_t = 3.2 \cdot 10^{-5}$, $10^{-6}$, and $3.2 \cdot 10^{-8}$. The volumetric strain step $\varepsilon_V$ was chosen to be -0.036, -0.020, and -0.011 for the respective target energies. Following consolidation selected configurations were expanded to estimate jamming thresholds. Figure 3 displays the resulting dependence of jamming threshold on target energy with the same initial positions for each value of $E_t$. Increased target energy enables configurations to access states with high jamming threshold. Hereafter $E_t = 3.2 \cdot 10^{-5}$ is utilized to obtain consolidated configurations having high jamming thresholds.

## A. Configurational distributions

To study the distribution of jamming thresholds for small assemblies of Platonic solids, 11 random initial configurations were consolidated for each shape. Selected configurations were then expanded in order to find the jamming threshold; the jamming thresholds of remaining configurations were determined by extrapolation of energy with the scaling behavior of other configurations in the ensemble. Phases form with distinct scaling of energy with respect to density; evidence and discussion of which will follow in the next section. The resulting jamming threshold distributions are shown in Fig. 4 grouped according to these phases. Tetrahedra, icosahedra, and octahedra exhibit uni-modal distributions, while dodecahedra and cubes display bi-modal and tri-modal distributions, respectively. In particular dodecahedra exhibited a 60 % probability of crystallizing at $\phi_J = 0.838$. Though cubes did not display crystallization, highly ordered layer structures formed for 50 % of the configurations at $\phi_J = 0.926$. The densities of these ordered structures are lower than the expected optimal packing densities due to the insufficient number of particles in the assembly. Such layered structures are similar to the irregular square tessellations proposed by Torquato and Jiao [5] but with vacancy defects due to finite system size. At intermediate density, cubes form another phase which is primarily distinguished by its mechanical properties and slight orientational disorder resulting from edge-face contacts.

Statistical properties of small system phases are listed in Table I. In addition, the particle sphericity $\psi$ listed is defined as the ratio of surface area between a sphere and a particle of the same volume (see [35]). With the exception of cubes, average jamming thresholds increase with sphericity for these small systems. As presented in the next section, this trend does not persist as system size increases. The jammed densities of disordered packings are substantially lower than the optimal densities recently reported in the literature. In contrast the average density for tetrahedra is reasonable when compared to the experimental density obtained for tetrahedral dice. Considering that tetrahedral dice fill 16 % more volume than inscribed tetrahedra, the inscribed tetrahedra of the experimental packing of Jaoshvili *et al.* [16] would have a density of 0.64 ± 0.02, which is very close to the average here. With such a substantial difference between optimal and jammed densities it is clear that optimal packing is improbable for athermal grains.

**B. Energy scaling**

Shown in Fig. 5a is the variation in energy with density for selected samples from the configurational distributions presented. Figures 5b and 5c reveal the scaling of the different phases observed. Samples in each phase group possess scaling exponents ranging over the values listed in Table II. Phases lacking order generally exhibited a nominal scaling exponent of 6 with respect to deviation from the jamming threshold that we denote as excess density ($\Delta\phi = \phi - \phi_J$), *i.e.* $E \sim (\Delta\phi)^6$. Such scaling is considered very soft when compared to the scaling of sphere packings. Among small ensembles, crystallization of dodecahedra and layering of cubes were observed, and, as a result, energy scaling displayed a nominal scaling exponent of 2 ($E \sim (\Delta\phi)^2$) with respect to excess density, distinct from that of disordered phases. The 'cubes edge' structure previously referred to exhibits a differing scaling exponent than layered cubes. This phase possessed sufficient density of edge-face contacts for mechanical percolation of the edge-face contact network to occur. As a result, these configurations exhibit an energy scaling exponent of 4 ($E \sim (\Delta\phi)^4$). As indicated in Fig. 5c, scaling of the cubes edge structure energy becomes softer as the jamming point is approached. We attribute this softening to decreased numerical accuracy at low $\Delta\phi$. As will be shown later, these highly ordered phases are suppressed as system size increases.

These scaling exponents can be understood in terms of the energy scaling of discrete contact topologies. A faceted particle possesses discrete geometric features (vertices, edges, and faces) that intersect the features of a contacting particle. Distinct combinations of these intersecting features we refer to as 'contact topologies'. The contact topology hierarchy possesses three levels with increasing degree of restraint between contacting particles – (1) vertex-face and edge-edge contacts, (2) edge-face contacts, and (3) face-face contacts. Within the hierarchy contact intersection volume varies with contact depth $d$ as $V \sim d^n$; levels 1, 2, and 3 of the hierarchy possess volume scaling exponents of $n$ = 3, 2, and 1, respectively. Within the contact mechanics model employed, energy scales as $E \sim V^m$ and therefore $E \sim d^{nm}$. For affine deformation, $\Delta\phi \sim d$ and therefore $E \sim (\Delta\phi)^{nm}$. Thus by utilizing $m$ = 2 in all simulations here, the nominal scaling exponents of 6, 4, and 2 are observed for the configurations exhibiting mechanical percolation of contacts with level 1, 2, and 3, respectively. This percolation phenomenon is unique to systems of soft, faceted particles and is essential to the scaling of mechanical properties in these systems. Scalar transport properties, such as effective thermal and electrical conductivity, of these systems will also display topologically dependent percolation physics.

## C. System size dependence and ordering

The number of particles was varied to determine limiting structures for each particle system with increasing degrees of freedom. For systems of 100 and 400 particles convergence tolerances of $D_E^{(k)}$ = $2.5 \cdot 10^{-13}$ and $6.25 \cdot 10^{-14}$ were respectively employed for consolidation. Coarser convergence tolerances during expansion of $10^{-8}$ were utilized for large systems and jamming threshold was approximated by extrapolation of average contact depth. Shown in Fig. 6, jamming threshold stabilizes as the number of particles increases. Also no crystallization was observed for configurations with 100 or 400 particles. Thus, the high probability of crystallization for dodecahedra in 25 particle configurations was enabled by the small system size. Therefore, it is desirable to understand the structural mechanisms behind order frustration, as well as a precise description of the disordered sub-structures present in these packings.

As discussed in the previous sections, disordered configurations of Platonic solids produce drastically lower jamming threshold than their ordered counterparts. Also, the scaling of energy with excess density changes dramatically between ordered and disordered phases. One expects the underlying structure of disordered and ordered configurations to be significantly different. Figure 7 compares such configurations at densities just above the jamming point ($0 < \Delta\phi < 0.02$). Crystallization is visually apparent for ordered dodecahedra, while its disordered counterpart possesses significant orientational and translational disorder. In contrast, only a few particles with orientational disorder disturb the translational order of cube systems. Specifically, only slight orientational disorder in the percolating edge-face cube system is necessary to induce structures that exhibit drastically different mechanical scaling.

Differences in these structural phases can be understood in terms of nematic ordering. We therefore calculate the nematic order parameter $S$ for configurations containing cubes, as described in Appendix E. This parameter approaches unity for highly oriented nematic phases. Configurations of layered cubes with scaling exponents of $\beta \sim 2$ exhibit the highest nematic order among all phases of cubes with $1.0000 \geq S > 0.9999$. Cubes edge structures with scaling exponents of $\beta \sim 4$ exhibit lesser nematic order with $0.999 > S > 0.99$. Finally, disordered structures with scaling exponents of $\beta \sim 6$ exhibit $S < 0.99$ with an average value of $S = 0.96$, displaying lowest nematic order. Thus only slight orientational disorder induces drastic contrast in the scaling of mechanical properties.

Figure 8 contains representations of the largest stable systems of particles investigated near the target energy $E_t = 3.2 \cdot 10^{-5}$. Disordered structures are apparent for large systems of tetrahedra, icosahedra, dodecahedra, and octahedra. Such visual evidence is consistent with translational and orientational correlation functions presented subsequently in Figs. 9, 10, and 11. In contrast, cubes exhibit mild nematic ordering along the dominant nematic director vector shown in the figure. Similar phases have been observed for thermal cuboids [36], superballs [37], and superellipsoids [38], but this is the first such study to confirm such a phase for athermal cubes. In contrast to small ordered systems, large systems of cubes exhibit substantially lower nematic order parameters of 0.93 and 0.87 for 100 and 400 cubes, respectively. Though the octahedron and cube form a dual pair and both possess octahedral symmetry, we find that octahedra tend to exhibit nematic order parameters of ~0.6, indicating little nematic order. Such behavior is in contrast with the nematic phase formed prior to crystallization of octahedra via Monte Carlo simulation (see [5]). This finding reveals that the triaxial symmetry of cube faces, rather than the triaxial symmetry of octahedron vertices, results in nematic order at the jamming point.

Radial distribution functions (RDFs) of large periodic systems presented in Fig. 9 were calculated by averaging over all particles and normalizing by the number of particles in an ideal gas volume element of the same density $\phi$, as is conventional practice (see [39]). As a result of homogeneous correlation, each RDF approaches unity at large radii. All RDFs are presented in terms of radius $r$ normalized by the nearest distance between contacting polyhedra $R_{min}$, and were calculated for configurations with energy near $E_t = 3.2 \cdot 10^{-5}$. At this minimum radius, all shapes display strong peaks as a result of face-face contacts. The RDFs for each particle shape vary dramatically, and are therefore grouped with RDFs displaying similar features. Firstly, the RDFs of dodecahedra and icosahedra, displayed in Fig. 9a, exhibit short-range translational order with features resembling those of the RDF for randomly packed spheres. The RDFs of these highly spherical faceted particles exhibit correlation peaks at positions similar to those of spheres. In particular, these dual polyhedra share a distinguishing feature with sphere RDFs differing from those of other shapes presented – closely-grouped secondary and tertiary peaks. Finally, peak intensities are reduced to 20 % above the mean value at a radius of $3.5R_{min}$ consistent with spheres. Subsequent peaks are offset from multiples of $R_{min}$.

Packings of the self-dual tetrahedron exhibit a RDF, shown in Fig. 9b, with ultra-low translational order dissimilar to the other Platonic solids. Such behavior is consistent with that observed experimentally for tetrahedral dice (see [16]), as displayed. RDFs of both shapes exhibit a peak between $R_{min}$ and $2R_{min}$ decaying to values less than 20% above the mean value at $2R_{min}$. Strong decay of the RDF of this disordered system contrasts with that of the higher-density thermal glasses generated by Haji-Akbari *et al.* [8] via thermodynamic Monte Carlo simulation. Finally, the dual pair of cubes and octahedra exhibit similar RDFs (Fig. 9c) with peaks at integer multiples of $R_{min}$. Peaks of the octahedra packing are broadened relative to those of cubes due to the lack of nematic order. Cubes and octahedra exhibit peaks with intensity greater than 20% above the mean up to $4R_{min}$ and $3R_{min}$, respectively.

Through visual observation of particle packing and orientational order parameter analysis, we have shown that large cube systems possess moderate nematic ordering. The influence of this nematic ordering is not clear from the calculation of RDF, because it exhibits diminishing correlation with radius. For nematic systems it is more appropriate to analyze the density correlation function with respect to coordinates longitudinal and transverse to the dominant nematic axis. These measures of translational correlation anisotropy have been utilized to distinguish between nematic and columnar phases of cut spheres [41] and are plotted in Fig. 10. The values shown were biased and normalized relative the mean values for the respective function. The fluctuations in both functions are very small relative to the mean values; intensity of the longitudinal function does not diminish with radius. In contrast, the transverse function exhibits strong correlation peaks which diminish with radius. Thus, translational order is present along the direction possessing nematic order.

In addition to the translational extent of ordering, the extent of orientational ordering is also interesting to assess. In particular we calculate the face-face angular correlation function $F(r)$ displayed in Fig. 11, according to the procedure described in [16], by averaging over all particles in large periodic systems. Calculation details are described in Appendix E. $F(r)$ measures the average anti-alignment of face normal vectors on particle surfaces. Aligned face-face contacts exhibit $F(r) = -1$, and $F(r)$ increases with decreasing alignment of faces. In the vicinity of $r \sim R_{min}$ the angular correlation function for each shape exhibits high face-face anti-alignment. Octahedra, dodecahedra, and icosahedra exhibit step-like $F(r)$ immediately following $R_{min}$. Thus, these particle systems exhibit nearly homogeneous orientational correlation, confirming orientational disorder. The nematic order of cubes is also reflected in their angular correlation function exhibiting multiple peaks. Tetrahedra exhibit orientational correlation diminishing to a constant value only after $3R_{min}$ in agreement with recent experimental results.

**D. Isostaticity and jamming contact number**

The variation of contact number with respect to excess density for sphere packings is known to exhibit square-root scaling for sphere systems independent of contact interaction model due to the tremendous degree of non-affine motion near the jamming point [3, 20]. We seek to investigate this dependence for packings of non-spherical Platonic solids. To perform this study, the average contact number during expansion from the high-energy consolidated state was calculated, as shown in Fig. 12. A single sample for each shape was chosen from the 25 particle ensemble with jamming threshold and radial distribution function similar to that of large systems for the same shape. Due to the high nematic order in small systems of cubes relative to large systems, they have been omitted from this analysis. As the jamming point is approached, these systems become ill-conditioned, and the number of iterations required to relax structures diverges. Therefore, average contact number could only be studied over a limited range of excess density. Over this limited range square-root scaling does appear to be consistent with the variation of average contact number. Statistical fluctuations relative to the curve fit are present that would likely be suppressed for larger systems.

Finally, square-root curve fits are used to extrapolate contact number to the jamming point. In Fig. 13 the jamming contact number $Z_J$ is plotted as a function of shape sphericity. The isostatic conjecture purports that the jamming contact number of three-dimensional particles $Z_{J,iso}$ depends on the number of particles $N$ and the number of degrees of freedom per particle $n$:

$$Z_{J,iso} = \frac{2nN - 6}{N} \tag{15}$$

For spheres with continuous rotational symmetries, $n = 3$ and $Z_{J,iso} = 5.76$ for a system of 25 particles. In contrast, for the Platonic solids studied here, continuous rotational symmetry is broken, and as a result $n = 6$ and $Z_{J,iso} = 11.76$. The actual calculated values of $Z_J$ are much lower than the value expected from the broken rotational symmetry of these shapes. Instead, values between the isostatic conjecture predictions are found for the shapes. Shapes possessing high sphericity (icosahedra and dodecahedra) display $Z_J$ just above the isostatic value for spheres with average contact number decreasing monotonically with sphericity $\psi$. The value of 8.6 ± 0.1 for tetrahedra is larger than the value of 6.3 ± 0.5 measured for a large system of tetrahedral dice [16]. The difference in average contact number is likely due to the high asphericity of tetrahedra relative to tetrahedral dice.

### IV. Conclusions

An energy-based approach to modeling the mechanical behavior of non-smooth particles has been implemented and utilized to study jamming of frictionless Platonic solids. The method does not incorporate thermal fluctuations as in the thermodynamic Monte Carlo approaches utilized to find optimal densities of non-smooth particles. For small particle systems, average jamming thresholds were obtained through configurational ensembles, and the resulting densities were substantially less than the previously reported optimal densities for each shape. In particular, our simulations produce tetrahedra with a jamming threshold consistent with experimental results. For small particle systems, dodecahedra can crystallize and cubes can order into layered structures with finite probability. These ordered structures are similar to the optimal structures previously predicted, but their formation is suppressed with increased system size. No prior reports have indicated such behavior, which is critical to the understanding of granular materials with non-smooth particle surfaces.

Radial distribution functions of the jammed structures were examined to quantify the extent of ordering in large configurations. The common symmetry of dual polyhedra results in similarity of radial distribution functions. The length scale of ordering for icosahedra, dodecahedra, and octahedra is approximately $3.5R_{min}$; also, icosahedra and dodecahedra display similar local structure to spheres. Tetrahedra exhibit order only over a length scale of $2R_{min}$, with a radial distribution function similar to that observed in recent experiments. The short-range orientational correlation of tetrahedra is consistent with experimentally measured correlation. In contrast, cubes exhibit much longer range order up to a length scale of $4R_{min}$. Large systems of cubes exhibit nematic ordering along a particular axis, despite lack of long-range translational order. Local ordering or the lack of it is crucial to description of these systems, and correlating these structures to macroscopic properties will be a subject of future investigation.

Aside from structural evidence based on the radial distribution function, ordered phases displayed power law scaling exponents for energy of 2 and 4 versus a scaling exponent of 6 for disordered phases. These effects are all linked to percolation of orientational order of faceted particles possessing discrete rotational symmetry. As the dramatic differences in mechanical properties resultant from topological contact networks affect mechanical properties, topological contact networks are expected to affect thermal and electrical transport properties within these microstructures. This topologically-dependent scaling phenomena is absent from assemblies of smooth particles, and this is the first such observance of this behavior.

Though all the Platonic solids possess broken continuous rotational symmetry, all possess contact numbers less than the isostatic value for asymmetric particles and are, hence, hypostatic. As evidenced by the close agreement of contact number for icosahedra and dodecahedra with the sphere isostatic value, the influence of rotational degrees of freedom is substantially less for these particles. Conversely, the remaining shapes studied display strong departure from the sphere isostatic value, and the average contact number at jamming correlates monotonically with shape sphericity. Prior work has revealed the hypostatic nature of smooth non-spherical particles [19] and tetrahedra [16], but ours is the first to correlate the degree of hypostaticity with sphericity of non-smooth particles.

The methods utilized herein provide a platform for further mechanical analysis of jamming of non-smooth particles. The findings also help to provide a more comprehensive picture of jamming phenomena across particle morphologies. Further study of the structural and local ordering phenomena in packings of faceted particles will aid development of granular materials possessing microstructures tailored for applications.

**Acknowledgements**

The authors acknowledge the Indo-US Science and Technology Forum for supporting Purdue-JNCASR exchanges through the Joint Networked Centre on Nanomaterials for Energy (Award 115-2008/2009-10) and the US National Science Foundation for workshop support (OISE-0808979) and a graduate research fellowship to KCS. KCS also thanks the Purdue Graduate School for financial support via the Charles C. Chappelle fellowship. The authors thank Damian Sheehy of The Mathworks for insightful correspondence regarding computational geometry, Adam Carlyle of the Rosen Center for Advanced Computing at Purdue for computing aid, Jayathi Murthy for utilization of computing resources, Andriy Kyrylyuk and Gary Delaney for valuable discussion regarding jamming phenomena, Mitch Mailman and Carl Shreck for insightful discussion about numerical structural optimization of granular materials, and Salvatore Torquato and Yang Jiao for helpful, encouraging comments.

# Appendices

## A. Cumulative quaternion-based rotations

Particles are numerically rotated from initial positions after each compression step to prevent accumulation of round-off error due to sequential rotation operator application. To facilitate rotation, the cumulative rotation is stored as a quaternion vector $\mathbf{q}_0$, where

$$\mathbf{q}_0 = [\mathbf{v}, s] = \left[\left(\sin\left(\frac{\theta}{2}\right)\right)\hat{\theta}, \cos\left(\frac{\theta}{2}\right)\right] \tag{16}$$

Here $s$ and $\mathbf{v}$ are the scalar and vector components of the quaternion, respectively. Incremental rotational trajectories determined during a conjugate gradient iteration are then converted to a quaternion vector $\mathbf{q}_1$. The angle of rotation $\theta$ about the unit vector axis $\hat{\theta}$ is determined from the rotational component of the conjugate gradient trajectory $\mathbf{s}_r^{(k)}$, where

$$\mathbf{s}_r^{(k)} = \theta\hat{\theta} \tag{17}$$

The resultant rotation direction is determined via quaternion multiplication, $\mathbf{q}_1\mathbf{q}_0$:

$$\mathbf{q}_1\mathbf{q}_0 = [s_1\mathbf{v}_0 + s_0\mathbf{v}_1 + \mathbf{v}_1 \times \mathbf{v}_0, s_1 s_0 - \mathbf{v}_1 \cdot \mathbf{v}_0] \tag{18}$$

The rotation operator is then applied to the coordinates of the particle for computation of intersection with other particles. This approach helps to prevent numerical distortion of particles and to maintain stability.

## B. Line search methods and termination criteria

A line search method that explicitly minimizes energy was utilized along each conjugate gradient search direction; the method employs combinations of golden searches and quadratic interpolations (see [42, 43]). The first Goldstein condition and a two-sided slope test (see [27]) were incorporated to assess convergence:

$$f(\alpha) - f(0) \leq \alpha \rho f'(0) \tag{19}$$

$$|f'(\alpha)| \leq -\sigma f'(0) \tag{20}$$

Here $\alpha$ and $f'(\alpha)$ are the position along the search direction and the projection of the gradient of energy along the search direction. Eq. 20 tends to be the more restrictive condition and is essential to finding an accurate minimum. Eq. 19 essentially ensures that $\alpha$ is not a maximum. In practice, a value $\rho = 0$ was required to achieve adequate computational efficiency, and since $\sigma = 0.033$ resulted in satisfactory convergence rates, this value was used in all simulations. Upon each line search we initially guess that the minimum is contained in the region of $\alpha$ satisfying:

$$0 < \alpha \leq 2 E^{(k-1)} D_E^{(k-1)} / f'(0) \tag{21}$$

Here $D_E^{(k-1)}$ is the relative energy after the previous conjugate gradient iteration. If energy calculated at the upper bound of the bracket does not satisfy Eq. 19, then the search interval is restricted in multiples of ten until it is satisfied. After this restriction, it is possible that the minimum is not contained within the bracket. Therefore, the search interval is then expanded in multiples of two until $f'(\alpha) > 0$ is satisfied. It is possible though that the initial interval satisfies Eq. 19 but does not in fact bracket the minimum. When $f'(\alpha) < 0$ this situation occurs and the search interval is correspondingly expanded in multiples of two until satisfied. These bracketing procedures are essential to convergence of the line search procedure. With these parameters, 2 bracketing function evaluations with 4 line search function evaluations are often required per conjugate gradient iteration.

## C. Intersection volume computation

Computation of intersections among polyhedra is essential to the structural optimization methods described. Overlap volume can be calculated with ease analytically for simple contact types such as vertex-face and edge-edge contacts, but for more complex types of contacts, analytical determination is very difficult to generalize. For instance, the drastic differences in contact topology require that generalized methods for determining intersection for arbitrary contacts are utilized. Methods from computational geometry are employed to generalize and stabilize such calculations.

In our implementation, candidate intersections between polyhedral particles are screened by performing bounding sphere contact detection (see [44]). Vertices on intersection volumes are found by determining intersections of the edges of both particles with the opposing particle via the ray intersection algorithm of Haines [45]. Convexity of the particles ensures that the intersection of the two particles is in fact convex. We therefore compute the Delaunay triangulation of the intersection points and calculate its volume through the resulting triangulated representation of the intersection geometry. Faces belonging originally to only one particle participating in the contact are identified as well. Triangulated representations of those faces are then utilized to calculate areas and centroids required for force and moment computations. Because volume and area of the intersection geometry are calculated with floating point arithmetic, volume and area of each simplex are sorted in ascending order prior to summation in order to minimize round-off error. This general approach to determining contact geometry without contact planes does not require explicit declaration of contact topology.

### D. Jamming threshold extrapolation

After obtaining jamming configurations near the target energy, configurations are expanded toward the jamming point. As a result of floating-point precision employed in our simulations, equilibrium configurations can only be simulated to finite values of $\Delta\phi$. This is a result of higher order contact topologies (face-face and edge-face contacts) being in equilibrium with lower order contact topologies (vertex-face and edge-face contacts).

Therefore, to accurately estimate the jamming threshold extrapolation techniques must be employed. As a result of affine motion during expansion toward the jamming point, we expect average contact depth to scale linearly with excess density, *i.e.* $d \sim \Delta\phi$. Numerically, we calculate depth for each contact as the ratio of contact volume to normal-projected surface area. We find the jamming threshold for a particular configuration by calculating the least-squares linear intercept for $\Delta\phi$ as a function of $d$.

### E. Nematic order and angular correlation computations

Nematic order parameters were calculated in order to assess uniaxial and biaxial ordering of cube and octahedron systems as in [36-38]. To identify the dominant nematic director vector for the system, orientational direction vectors for each particle must be grouped into sets having similar direction. For cubes, these direction vectors correspond to face unit normals, while for octahedra they correspond to vertex unit position vectors relative to particle centroids. To do this, we choose one particle's axes as a reference and match axes of the other particles to those reference axes. From this procedure three sets of aligned axes are obtained – $\{\hat{u}_i\}$, $\{\hat{v}_i\}$, and $\{\hat{w}_i\}$. For axes set $\{\hat{u}_i\}$ we calculate the nematic tensor $\mathbf{Q}^{uu}$ as a sum over all particles [36]:

$$\mathbf{Q}_{\alpha\beta}^{uu} = N^{-1}\sum_{i=1}^{N}\left(\frac{3}{2}u_{i\alpha}u_{i\beta} - \frac{1}{2}\delta_{\alpha\beta}\right) \qquad (22)$$

We calculate $\mathbf{Q}^{vv}$ and $\mathbf{Q}^{ww}$ for axes sets $\{\hat{v}_i\}$ and $\{\hat{w}_i\}$ as well. For each nematic tensor we determine the dominant eigenvalues and eigenvectors and assign the maximal eigenvalue among those three sets as the uniaxial nematic order parameter $S$. The nematic director vector is then assigned as the eigenvector corresponding to the maximal eigenvalue among the three axes sets.

Face-face angular correlation functions were calculated according to the approach described by Jaoshvili *et al.* [16] to quantify orientational alignment. The face-face angular correlation between particles $q$ and $l$ is given as:

$$F_{ql} = \min_{\langle ij \rangle}\left(\hat{n}_{qi}\cdot\hat{n}_{lj}\right) \qquad (23)$$

where $i$ and $j$ represent the set of all faces of particles $q$ and $l$, respectively. $\hat{n}$ represents the unit normal vector of a given face. $\langle ij \rangle$ represents the set of all possible combinations of $i$ and $j$. The face-face angular correlation function is determined as:

$$F(r) = \left\langle F_{ql}\delta\left(r - |\mathbf{r}_q - \mathbf{r}_l|\right)\right\rangle \qquad (24)$$

where $\delta(r)$ is the Dirac delta function. To numerically evaluate this function we compute $F_{ql}$ for all particle combinations of $q$ and $l$. The set of all $F_{ql}$ are then binned with respect to $|\mathbf{r}_q - \mathbf{r}_l|$, and the average value for the bin at radius $r$ is assigned to $F(r)$.

**Tables**

**Table I** – Average jamming threshold for jammed assemblies of 25 Platonic solids determined via 11 random initial configurations. Uncertainties shown represent the 95 % confidence intervals for each spectrum.

|                     | $\psi$ | $\phi_J$          | $\phi_{max}$   |
|---------------------|--------|-------------------|----------------|
| tetrahedra          | 0.671  | $0.611 \pm 0.037$ | $0.856^{(a)}$  |
| octahedra           | 0.846  | $0.677 \pm 0.011$ | $0.947^{(b)}$  |
| dodecahedra         | 0.910  | $0.684 \pm 0.009$ | $0.904^{(b)}$  |
| icosahedra          | 0.939  | $0.727 \pm 0.029$ | $0.836^{(b)}$  |
| cubes               | 0.806  | $0.773 \pm 0.057$ | 1.000          |
| cubes edge          | 0.806  | $0.796 \pm 0.051$ | 1.000          |
| dodecahedra crystal | 0.910  | 0.838             | $0.904^{(b)}$  |
| layered cubes       | 0.806  | 0.926             | 1.000          |

[a] ref. 9
[b] ref. 5

**Table II** – Average energy scaling exponents for jammed assemblies of 25 Platonic solids. Uncertainties shown represent the maximal deviation of the sample exponents from the average.

|  | $\beta$ |
|---|---|
| disordered phases | |
| tetrahedra | $6.34 \pm 0.92$ |
| dodecahedra | $5.44 \pm 0.08$ |
| octahedra | $5.67 \pm 0.24$ |
| icosahedra | $5.47 \pm 0.58$ |
| cubes | $5.49 \pm 0.91$ |
| ordered phases | |
| cubes edge | $4.61 \pm 0.29$ |
| dodecahedra crystal | $2.22 \pm 0.22$ |
| layered cubes | $1.99 \pm 0.14$ |

**Figure captions**

**FIG. 1** – (color online) Platonic solids from left to right: tetrahedron, icosahedron, dodecahedron, octahedron, and cube.

**FIG. 2** – (color online) Average energy as a function of density for a jammed assembly of 25 cubes with $E_t$ = $3.2 \cdot 10^{-5}$ and $\varepsilon_V$ = -0.036 during consolidation and subsequent expansion toward the jamming threshold. The inset shows how average contact depth for the expansion process was linearly extrapolated to approximate the jamming threshold. Energy is normalized by $V_p^{1/3} Y$, where $V_p$ is particle volume.

**FIG. 3** – (color online) Jamming threshold as a function of target energy for configurations of 25 Platonic solids. 95 % confidence intervals based on extrapolation of contact depth are represented by error bars. Energy is normalized by $V_p^{1/3} Y$, where $V_p$ is particle volume.

**FIG. 4** – (color online) Configurational spectrum of jamming threshold for assemblies of 25 Platonic solids produced via 11 random initial configurations.

**FIG. 5** – (color online) Average energy for jammed assemblies of 25 Platonic solids. (a) average energy as a function of density. Average energy as a function of excess density for (b) disordered phases and (c) ordered phases. Energy is normalized by $V_p^{1/3} Y$, where $V_p$ is particle volume.

**FIG. 6** – (color online) Scaling of jamming threshold as a function of number of particles for jammed assemblies of 25 Platonic solids. 60 % confidence intervals are represented by error bars. Statistics for $N$=25 configurations were determined from the sample set of $\phi_J$ obtained for different initial configurations.

For $N$=100 and 400, $\phi_J$ for only one configuration is presented with uncertainty due to the standard error of contact depth extrapolation.

**FIG. 7** – (color online) Depictions of the ordered and disordered configurations of dodecahedra and cubes in the primary cubic cell with boundaries indicated in black. Highly ordered (a) crystallized dodecahedra, (c) layered cubes, and (d) semi-ordered cubes with percolating edge-face contacts. Highly disordered (b) dodecahedra and (e) cubes. Arrows indicate the dominant uniaxial nematic director vector for each phase of cubes.

**FIG. 8** – (color online) Jammed assemblies of 400 (a) tetrahedra, (b) icosahedra, (c) dodecahedra, (d) octahedra, and (e) cubes. (f) cubes are oriented clearly displaying nematic order in the direction of the arrow. Boundaries of primary cubic cells are indicated in black.

**FIG. 9** – (color online) Radial distribution functions for jammed assemblies of 400 (a) icosahedra, dodecahedra, and spheres, (b) tetrahedra, tetrahedral dice, and tetrahedra thermal glass, and (c) octahedra and cubes. The RDF of the tetrahedra thermal glass has been scaled for viewing purposes.

**FIG. 10** – Anisotropic radial distribution functions for a jammed assembly of 400 cubes. The longitudinal distribution function is calculated along the dominant direction of nematic order, while the transverse distribution function is calculated in the plane normal to that direction. The peak at $r/R_{min} < 1$ is a consequence of projecting transverse (longitudinal) neighbors onto the longitudinal (transverse) direction.

**FIG. 11** – (color online) Face-face angular correlation functions for jammed assemblies of 400 Platonic solids.

**FIG. 12** – (color online) Scaling of average contact number as a function of excess density for jammed assemblies of 25 Platonic solids. Curves represent square-root scaling fits.

**FIG. 13** – (color online) Jammed contact number as a function of particle sphericity for assemblies of 25 Platonic solids. Error bars were calculated based on the standard error for square-root extrapolation.

**Figure 1** (two-column format)

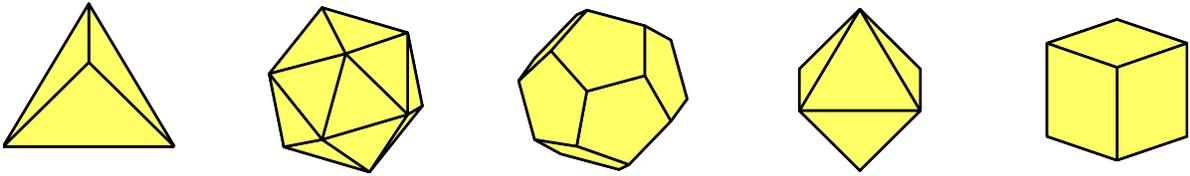

**Figure 2**

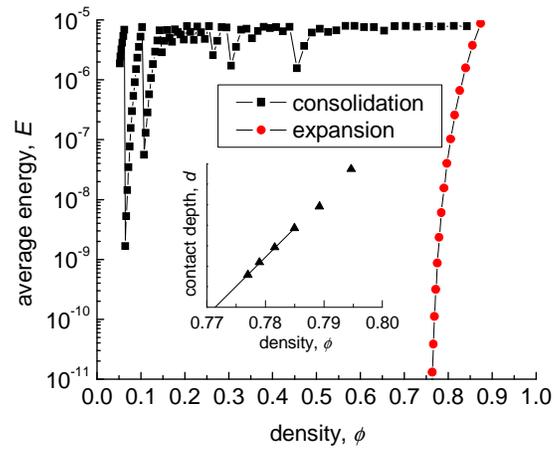

**Figure 3**

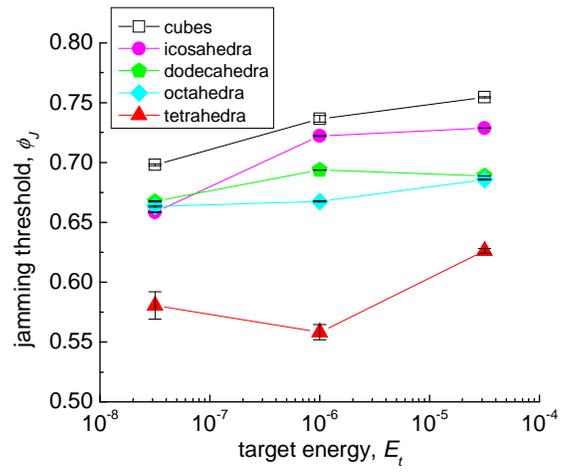

**Figure 4**

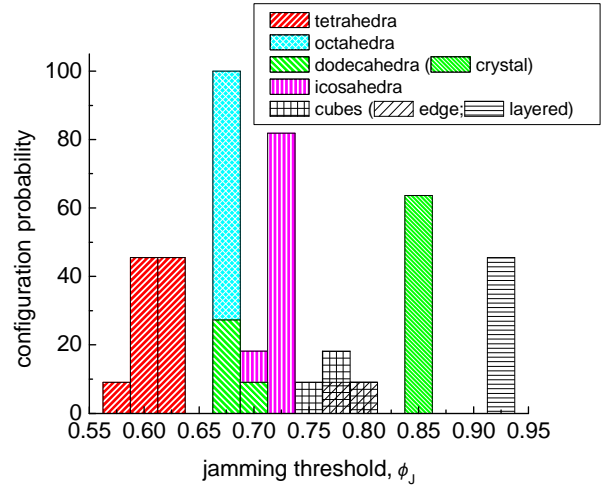

**Figure 5**

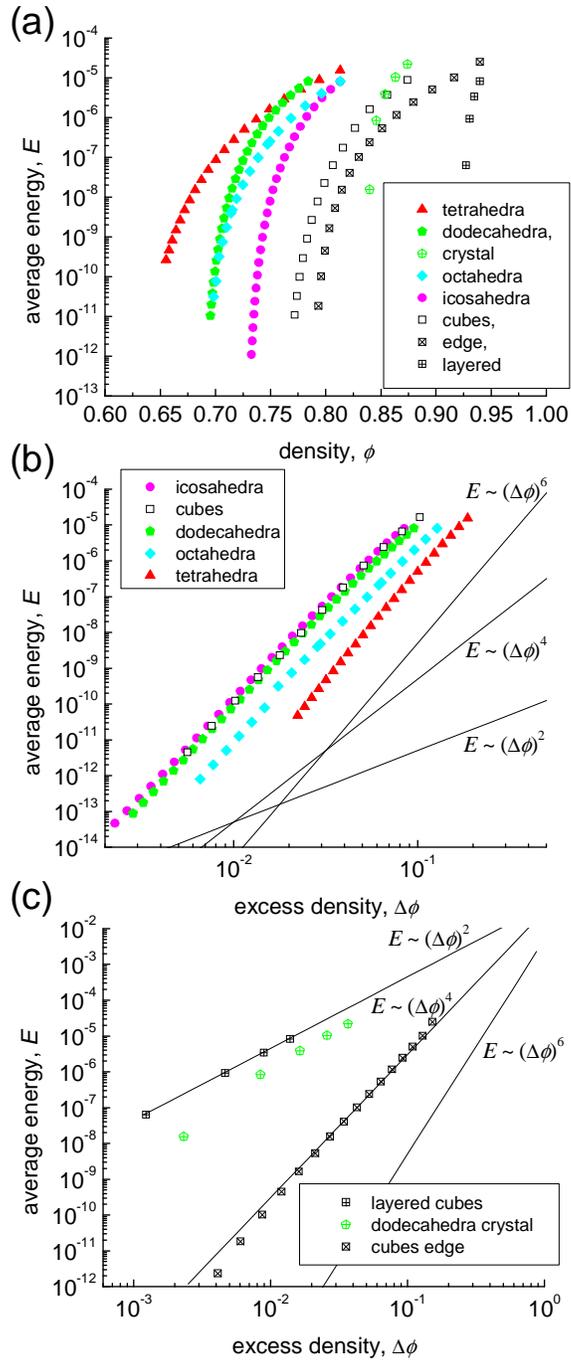

**Figure 6**

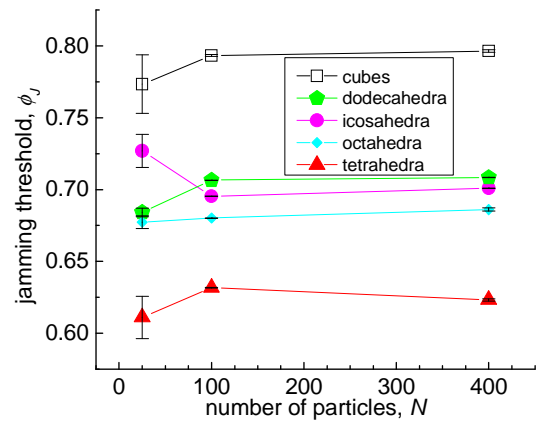

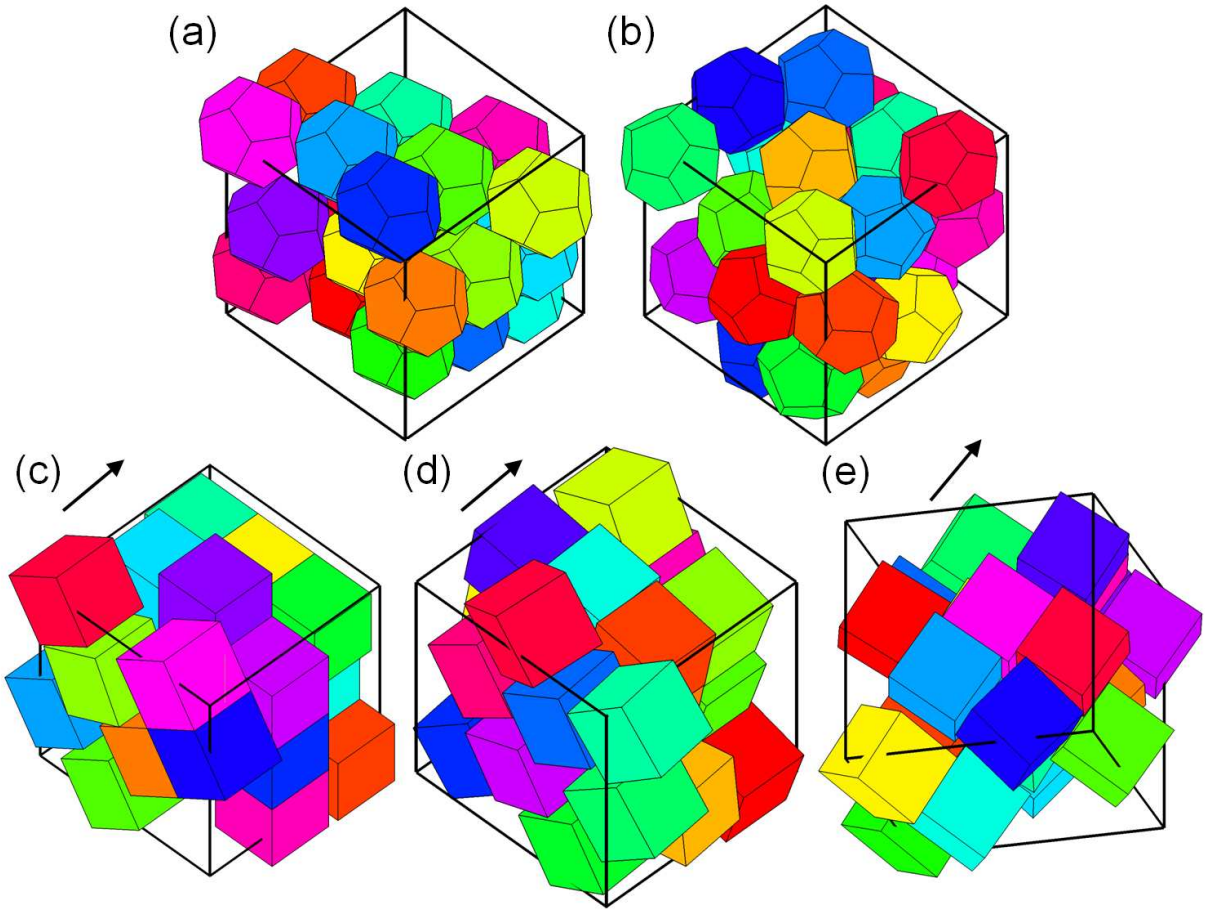

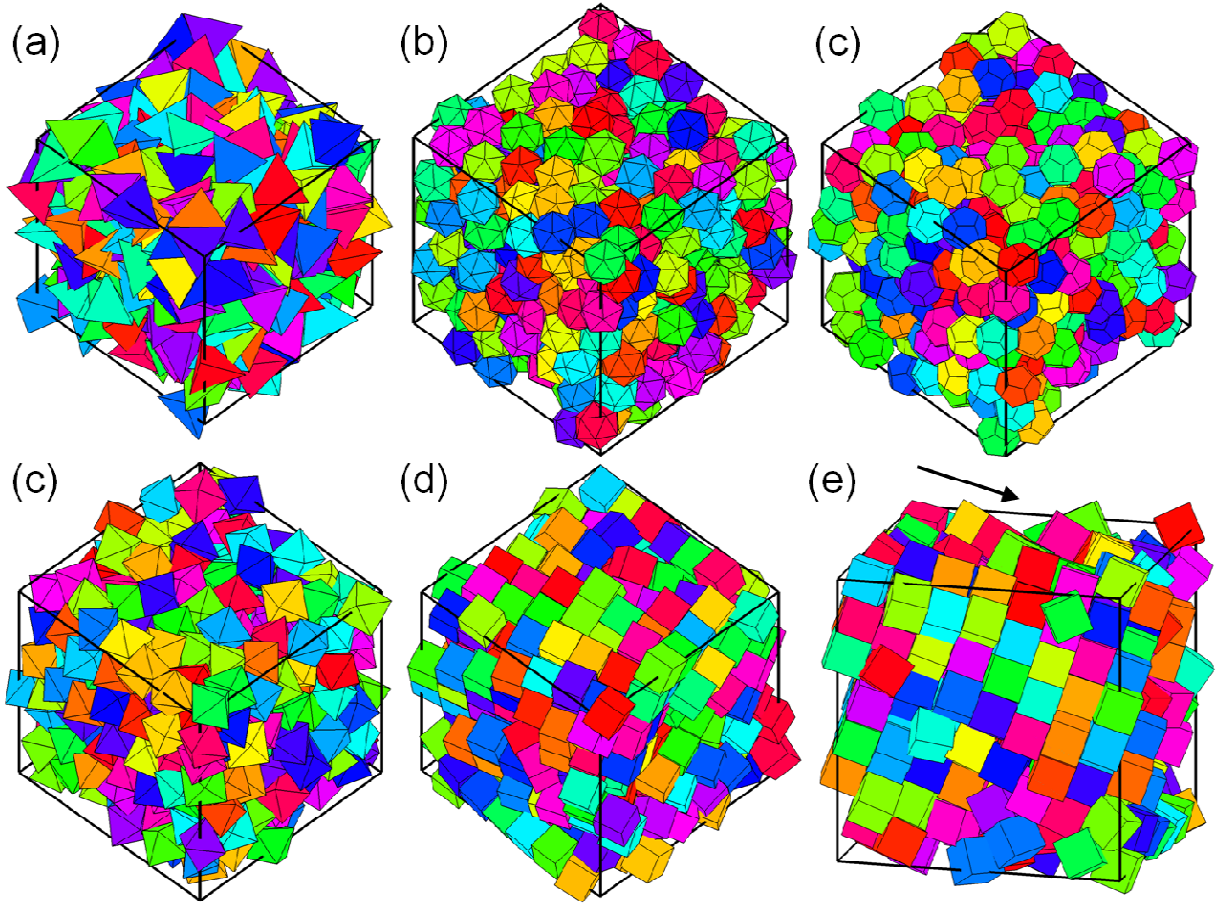

**Figure 9**

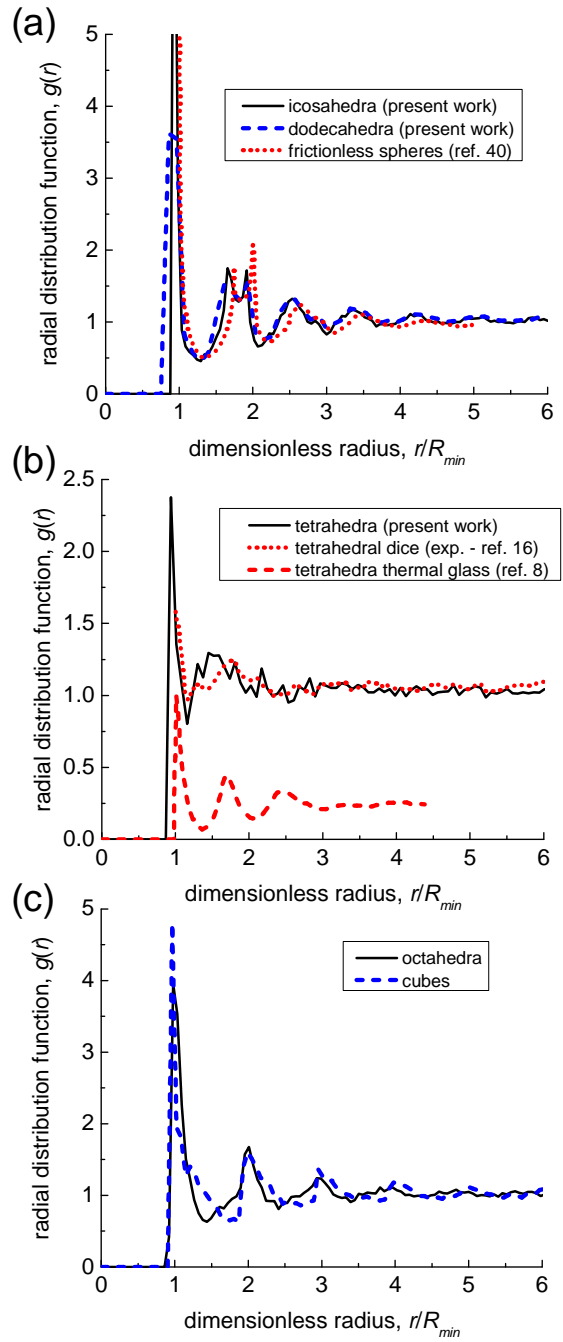

**Figure 10**

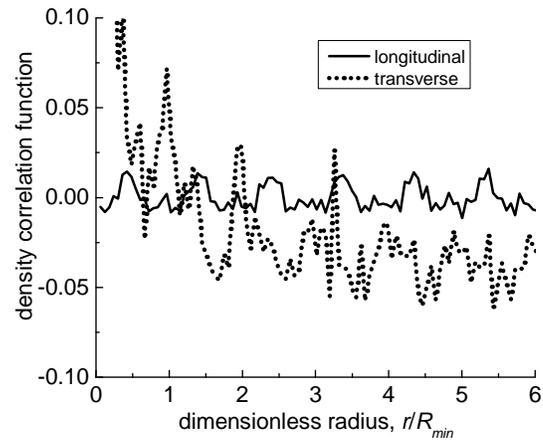

**Figure 11**

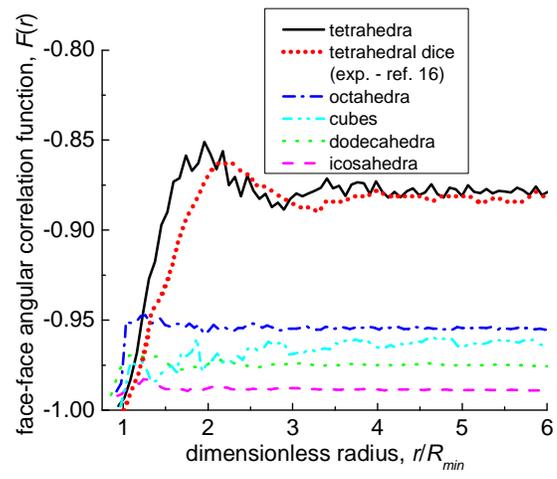

**Figure 12**

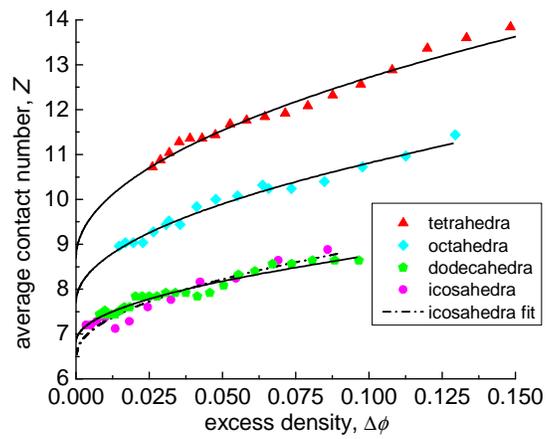

**Figure 13**

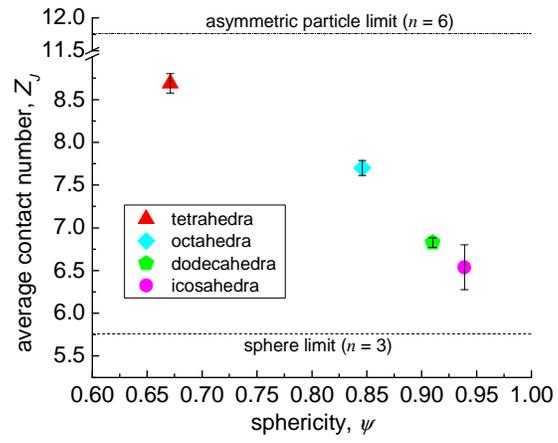

**References**

[1]     M. van Hecke, J. Phys.- Cond. Mat. **22**, 033101 (2010).

[2]     H. M. Jaeger, S. R. Nagel, and R. P. Behringer, Rev. Mod. Phys. **68**, 1259 (1996).

[3]     C. S. O'Hern, L. E. Silbert, A. J. Liu, and S. R. Nagel, Phys. Rev. E **68**, 011306 (2003).

[4]     T. C. Hales, Ann. Math. **162**, 1063 (2005).

[5]     S. Torquato, and Y. Jiao, Phys. Rev. E **80**, 041104 (2009).

[6]     S. Torquato, and Y. Jiao, Nature **460**, 876 (2009).

[7]     S. Torquato, and Y. Jiao, Phys. Rev. E **81**, 041310 (2010).

[8]     A. Haji-Akbari, M. Engel, A. S. Keys, X. Y. Zheng, R. G. Petschek, P. Palffy-Muhoray, S. and S. C. Glotzer, Nature **462**, 773 (2009).

[9]     E. R. Chen, M. Engel, and S. C. Glotzer, Discrete and Computational Geometry **44**, 253 (2010).

[10]    D. J. Hoylman, Bull. Am. Math. Soc. **76**, p. 135 (1970).

[11]    J. H. Conway, and S. Torquato, Proc. Nat. Acad. Sci. USA **103**, p. 10612 (2006).

[12]    E. R. Chen, Discrete and Computational Geometry **40**, pp. 214-240 (2008).

[13]    Y. Kallus, V. Elser, and S. Gravel, Discrete and Computational Geometry **44**, pp. 245-252 (2010).

[14]    A. Donev, F. H. Stillinger, P. M. Chaikin, and S. Torquato, Phys. Rev. Lett. **92**, 255506 (2004).

[15]    Y. Jiao, F. H. Stillinger, and S. Torquato, Phys. Rev. E **79**, 041309 (2009).

[16]    A. Jaoshvili, A. Esakia, M. Porrati, and P. M. Chaikin, Phys. Rev. Lett. **104**, 185501 (2010).

[17]    S. Torquato, T. M. Truskett, and P. G. Debenedetti, Phys. Rev. Lett. **84**, 2064 (2000).

[18]    M. Mailman, C. F. Schreck, C. S. O'Hern, and B. Chakraborty, Phys. Rev. Lett. **102**, 255501 (2009).

[19]   A. Donev, R. Connelly, F. H. Stillinger, and S. Torquato, Phys. Rev. E **75**, 051304 (2007).

[20]   W. G. Ellenbroek, E. Somfai, M. van Hecke, and W. van Saarloos, Phys. Rev. Lett. **97**, 258001 (2006).

[21]   S. Torquato, and F. Stillinger, Rev. Mod. Phys. **82**, pp. 2633-2672 (2010).

[22]   G. J. Gao, J. Blawzdziewicz, and C. S. O'Hern, Phys. Rev. E **74**, 061304 (2006).

[23]   The term 'collective jamming' was first invoked by Ref. [24] for hard spheres, and its equivalency to mechanical stable systems of soft spheres with fixed periodic boundaries has been discussed in Ref. [25-26].

[24]   S. Torquato and F. H. Stillinger, J. Phys. Chem. B **105**, 11849 (2001).

[25]   A. Donev, S. Torquato, F. H. Stillinger, and R. Connelly, Phys. Rev. E **70**, 043301 (2004).

[26]   C. S. O'Hern, L. E. Silbert, A. J. Liu, and S. R. Nagel, Phys. Rev. E **70**, 043302 (2004).

[27]   R. Fletcher, *Practical methods of optimization* (Wiley, Chichester, 1987), pp. 29, 80.

[28]   E. Polak, *Computational methods in optimization: a unified approach* (Academic Press, New York, 1971).

[29]   R. Fletcher, and C. M. Reeves, Computer Journal **7**, 149 (1964).

[30]   M. J. D. Powell, Mathematical Programming **11**, 42 (1976).

[31]   Y. T. Feng, K. Han, and D. R. J. Owen, in 3[rd] MIT Conference on Computational Fluid and Solid Mechanics, edited by K. J. Bathe (Cambridge, MA, 2006), pp. 210-214.

[32]   T. Pöschel and T. Schwager, *Computational granular dynamics: models and algorithms* (Springer-Verlag, Berlin, 2005), pp. 88-100.

[33]   Y. T. Feng and D. R. J. Owen, Engg. Comput. **21**, 265 (2004).

[34]   F. Alonso-Marroquin, S. Luding, H. J. Herrmann, and I. Vardoulakis, Phys. Rev. E **71**, 051304 (2005).

[35]    H. Wadell, J. Geology **43**, 250 (1935).

[36]    B. S. John, A. Stroock, and F. A. Escobedo, J. Chem. Phys. **120**, 9383 (2004).

[37]    R. D. Batten, F. H. Stillinger, and S. Torquato, Phys. Rev. E **81**, 061105 (2010).

[38]    G. W. Delaney and P. W. Cleary, EPL **89**, 34002 (2010).

[39]    M. P. Allen and D. J. Tildesley, *Computer simulation of liquids* (Clarendon Press, Oxford, 1987), pp. 183-184.

[40]    L. E. Silbert, D. Ertas, G. S. Grest, T.C. Halsey, and D. Levine, Phys. Rev. E **65**, 031304 (2002).

[41]    J. A. C. Veerman, and D. Frenkel, Phys. Rev. A **45**, 5632 (1992).

[42]    R. P. Brent, *Algorithms for minimization without derivatives* (Prentice-Hall, Englewood Cliffs, N.J., 1972), pp. 61-80.

[43]    G. E. Forsythe, M. A. Malcolm, and C. B. Moler, *Computer methods for mathematical computations* (Prentice-Hall, Englewood Cliffs, N.J., 1977), pp. 178-182.

[44]    C. Ericson, *Real-time collision detection* (Morgan Kaufmann Publishers, San Francisco, 2005), pp. 75-101.

[45]    E. Haines, in Graphics Gems II, edited by J. Arvo (Morgan Kaufmann Publishers, 1994), pp. 247-250.